\documentclass[twocolumn,epjc3]{svjour3}

\usepackage{graphicx}
\usepackage{bm}
\usepackage{amsmath}
\usepackage{amssymb}
\usepackage{hyperref}

\newcommand{\xbj}{x}
\newcommand{\zh}{z_h}
\newcommand{\nslash}{\kern 0.2 em n\kern -0.50em /}
\newcommand{\kslash}{\kern 0.2 em k\kern -0.45em /}
\newcommand{\lslash}{\kern 0.2 em l\kern -0.50em /}
\newcommand{\pslash}{\kern 0.2 em p\kern -0.50em /}
\newcommand{\Sslash}{\kern 0.2 em S\kern -0.50em /}
\newcommand{\Pslash}{\kern 0.2 em P\kern -0.50em /}
\newcommand{\Dslash}{\kern 0.2 em D\kern -0.65em /\kern 0.15em}

\newcommand{\bp}{\boldsymbol{p}_T}
\newcommand{\bP}{\boldsymbol{P}_T}
\newcommand{\bk}{\boldsymbol{k}_T}

\newcommand{\eps}{\epsilon}
\newcommand{\ssh}{\!\!\!/}
\newcommand{\Tr}{\operatorname*{Tr}\nolimits}

\newcommand{\ph}{\phi_h}

\journalname{Eur. Phys. J. C}

\begin{document}

\title{Beam Spin Asymmetries of Charged and Neutral Pion Production in Semi-inclusive DIS}

\author{Wenjuan Mao\thanksref{addr1}
      and
      Zhun Lu\thanksref{e1,addr1}}
\thankstext{e1}{email: zhunlu@seu.edu.cn}
\institute{Department of Physics, Southeast University, Nanjing 211189, China \label{addr1}}
\maketitle

\begin{abstract}
We present a study on the beam single spin asymmetries $A_{LU}^{\sin\phi_h}$ of $\pi^+$, $\pi^-$ and $\pi^0$ production in semi-inclusive deep inelastic scattering process, by considering Collins effect and the $g^\perp D_1$ term simultaneously.
We calculate the twist-3 distributions $e(x, \bm k_T^2)$ and $g^\perp(x,\bm k_T^2)$ for the valence quarks inside the proton in a spectator model.
We consider two different options for the form of diquark propagator, as well as two different choices for the model parameters in the calculation.
Using the model results, we estimate the beam spin asymmetries $A_{LU}^{\sin\phi_h}$ for the charged and neutral pions and compare the results with the measurement from the HERMES Collaboration. We also make predictions on the asymmetries at CLAS with a $5.5 \,\textrm{GeV}$ beam using the same model results.
It is found that different choices for the diquark propagator will not only lead to different expressions for the distribution functions, but also result in different sizes of the asymmetries. Our study also shows that, although the spectator model calculation can describe the asymmetries for certain pion production in some kinematic regions, it seems difficult to explain the asymmetries of pion production for all three pions in a consistent way from the current versions.
\end{abstract}

\section{Introduction}

The single spin asymmetry (SSA) appearing in high energy scattering process, known as a powerful tool to probe the internal structure of the nucleon, has attracted extensive attention in QCD spin physics \cite{bdr,D'Alesio:2007jt,Barone:2010ef,Boer:2011fh} in the past two decades.
Sizable SSAs in the semi-inclusive deep inelastic scattering (SIDIS) were measured by the HERMES Collaboration~\cite{hermes00,hermes01,hermes03,hermes05,hermes07,hermes09,hermes10}, the COMPASS Collaboration~\cite{compass05,compass06,compass10,Adolph:2012sn,Adolph:2012sp}, and the Jefferson Lab (JLab)~ \cite{clas04,clas10,Qian:2011py,Aghasyan:2011ha,Aghasyan:2011zz}.
As a particular case of SSAs, the beam spin asymmetry was observed in SIDIS by colliding the longitudinal polarized electron~\cite{clas04,Aghasyan:2011ha,Aghasyan:2011zz} or positron beam~\cite{hermes07} on the unpolarized nucleon target. Originally, two different mechanisms have been proposed to explain the observed asymmetry. One is the so called Boer-Mulders effect~\cite{Yuan:2004plb}, which suggests that the beam SSA is contributed by the convolution of the Boer-Mulders function $h_1^\perp$~\cite{Boer:1998prd} and the fragmentation function (FF) $E$~\cite{Yuan:2004plb,Gamberg:2003pz}.
The other refers to the Collins effect~\cite{Gamberg:2003pz,Efremov:2002ut}, which indicates that the asymmetry is resulted from the coupling of the distribution $e$~\cite{Jaffe:prl67,Jaffe:npb375} with the Collins FF $H_1^\perp$~\cite{Collins:1993npb}.

Apart from the above-mentioned mechanisms, a new source giving rise to the beam SSA at the twist-3 level has been found through model calculations~\cite{Afanasev:2003ze,Metz:2004epja22}.
This mechanism involves a new twist-3 transverse momentum dependent (TMD) distribution function (DF), denoted by $g^\perp$~\cite{Bacchetta:g2004plb}, which appears in the decomposition of the quark correlator if the dependence on the light-cone vector is included.
Furthermore, the function $g^\perp$ is time-reversal odd, thus it needs initial-state or final-state interaction~\cite{Brodsky:2002plb,Collins:2002plb,jy02} via soft gluon exchange to receive a nonzero result.
In this sense, $g^\perp$ is often viewed as an analog of the Sivers function~\cite{Sivers:1991prd,Anselmino:1998plb} at twist 3.
In Refs.~\cite{Afanasev:2006prd74,Gamberg:2006plb,Lu:2012plb}, a scalar diquark model had been adopted to calculate $g^\perp$.
In our recent work~\cite{wjmao:2012prd}, we extended the previous calculations by using a spectator model which includes the axial-vector diquark to obtain $g^\perp$ for both the $u$ and $d$ quarks. Moreover, by using the model results, we estimated the beam SSA in neutral pion production at CLAS and HERMES, and compared the calculations with the experimental data in that work. It was found that the $T$-odd twist-3 distribution $g^\perp$ may play an important role in the beam SSA in SIDIS~\cite{wjmao:2012prd}.

In this work, we study the contribution of $g^\perp$, as well as that of the Collins effect, to the beam SSAs for the pion production of all three flavors in SIDIS.
To this end we recalculate the twist-3 TMD DFs $g^\perp$ and $e$ in the spectator model, as shown in Sec.~\ref{formulation}.
Specifically, we consider two different options for the propagator of the axial-vector diquarks, as well as two different relations between quark flavors and diquark types for comparison.
The first choice we adopt is the one used in Ref.~\cite{wjmao:2012prd}, and it has been proposed previously in Ref.~\cite{Bacchetta:2008prd}.
For the second choice we apply the option used in Ref.~\cite{Bacchetta:plb578}.
In Section.~\ref{BSAs},
we present an analysis on the beam SSAs for the charged and neutral pion productions at the kinematics of HERMES~\cite{hermes07}, and compare the results with the experimental data, based on the DFs $e$ and $g^\perp$ for the $u$ and $d$ valence
quarks obtained from the two options.
For a further test, we also make the predictions on the beam SSAs for $\pi^+$, $\pi^-$ and $\pi^0$ at CLAS with the beam energy $E_e=5.5~\textrm{GeV}$.
We also compare the contributions from the Collins effect and DF $g^\perp$ in the numerical calculation.
Some conclusions are addressed in Sec.~\ref{conclusion}.

\section{Model calculation of TMD DFs $e$ and $g^\perp$ in the spectator model with an axial-vector diquark}
\label{formulation}

In this section, we present our calculation on the twist-3 TMD DFs $e$ and
$g^\perp$.
The DF $e$ has been calculated by several models, such as the spectator model \cite{Gamberg:2003pz,Jakob:1997npa}, the chiral quark-soliton model \cite{Schweitzer:2003prd67}, and the bag model \cite{Avakian:2010prd81}.
There was also an attempt to extract $e$~\cite{Efremov:2002ut} from SIDIS data~\cite{clas04}.
The DF $g^\perp$ has been studied by the spectator model~\cite{Afanasev:2006prd74,Gamberg:2006plb,Lu:2012plb,wjmao:2012prd}.

The gauge-invariant quark-quark correlator for the unpolarized nucleon can be expressed as
\begin{align}
\Phi^{[+]}(x,\bm k_T)&=\int {d\xi^- d^2\xi_T\over (2\pi)^3}e^{ik\cdot\xi}
\langle P|\bar{\psi}_j(0)\mathcal{L}[0^-,\infty^-]\nonumber\\
& \times \mathcal{L}[\bm 0_T,\bm \xi_T]\mathcal{L}[\infty^-,\xi^-]\psi_i(\xi)|P\rangle\,,
\label{Phi}
\end{align}
where $[+]$, corresponding to the SIDIS process, denotes that the gauge-link appearing in $\Phi$ is future-pointing;
$k$ and $P$ are the momenta of the active quark and the target nucleon, respectively. At the twist-3 level, the correlator (\ref{Phi}) can be decomposed into~\cite{Bacchetta:0611265}:
\begin{align}
\Phi^{[+]}(x,\bm k_T)\bigg{|}_{\textrm{twist-3}}
&= {M\over 2P^+}\left\{e-g^{\perp}\gamma_5{\epsilon_T^{\rho\sigma}\gamma_\rho k_{T\sigma}\over M}+\cdots\right\},
\end{align}
here $\cdots$ denotes the other twist-3 DFs that are not relevant in our calculation.
The TMD DFs $e$ and $g^\perp$ may be obtained from the correlator via the following traces:
\begin{align}
\frac{M}{P^+}e(x,\bm k_T^2)   & =
\frac{1}{2}\Tr[\Phi^{[+]}\gamma^{\alpha}],
 \label{e}\\
\frac{\eps_{T}^{\alpha\rho} k_{T \rho}^{}}{P^+} \,
  g^{\perp}(x,\bm{k}_{T}^{2})& =
-\frac{1}{2}\Tr[\Phi^{[+]}\gamma^{\alpha}\gamma_5]. \label{phitr2}
\end{align}

In the following, we will calculate $e$ and $g^\perp$ in the spectator model adopted in our previous work \cite{wjmao:2012prd}, which was originally developed in Ref.~\cite{Bacchetta:2008prd}.
For comparison, we will also consider the variations of the spectator model and calculate the same twist-3 TMD DFs using the version adopted in Ref.~\cite{Bacchetta:plb578}.

Using the spectator approximation, we can insert a complete set of intermediate states $|P-k\rangle$~\cite{Jakob:1997npa} into the correlator (\ref{Phi}), which has the following analytic form in the lowest order:
\begin{align}
\Phi^{(0)}(x,\bm k_T)=\frac{1}{(2\pi)^3}\,\frac{1}{2(1-x)P^+}\,
\overline{\mathcal{M}}^{(0)}\, \mathcal{M}^{(0)} ,
\label{eq:Phi-tree-spect}
\end{align}
where $\mathcal{M}^{(0)}$ is the nucleon-quark-spectator scattering amplitude at the tree level:
\begin{align}
\mathcal{M}^{(0)}
&=\langle P-k |\psi(0) |P\rangle \nonumber\\
&=\begin{cases}
  \displaystyle{\frac{i}{k\ssh-m}}\, \Upsilon_s \, U(P)\\
  \displaystyle{\frac{i}{k\ssh-m}}\, \varepsilon^*_{\mu}(P-k,\lambda_a)\,
        \Upsilon_v^{\mu}\, U(P),
  \end{cases}
\label{eq:m-tree}
\end{align}
and $\bar{\mathcal{M}}^{(0)}$ is its Hermitian conjugation.
Here $\Upsilon_{s/v}$ denotes the nucleon-quark-diquark vertex ($s$ for the scalar diquark and $v$ for the axial-vector diquark) and has the following form~\cite{Jakob:1997npa}
\begin{align}
\Upsilon_s(k^2) = g_s(k^2),~~~
\Upsilon_v^\mu(k^2)={g_v(k^2)\over \sqrt{2}}\gamma^\mu\gamma^5, \label{eq:vertex}
\end{align}
$\varepsilon_{\mu}(P-k,\lambda_a)$ is the polarization vector of the axial-vector diquark,
and
$g_X(k^2)$ is the form factor of the coupling. To regularize the light-cone divergent appearing in the calculation of $g^\perp$, we choose the dipolar form for $g_X(k^2)$:
\begin{align}
g_X(k^2)&= N_X {k^2-m^2\over |k^2-\Lambda_X^2|^2} \nonumber\\
&= N_X{(k^2-m^2)(1-x)^2\over
(\bm k_T^2+L_X^2)^2},~~ X=s,v, \label{eq:gx}
 \end{align}
where $\Lambda_X$ is the cutoff parameter, $N_X$ is the coupling constant, and
$L_X^2$ has the form
\begin{align}
L_X^2=(1-x)\Lambda_{X}^2 +x M_{X}^2-x(1-x)M^2.
\end{align}

Inserting Eqs. (\ref{eq:m-tree}), (\ref{eq:vertex}) and (\ref{eq:gx}) into Eq.~(\ref{eq:Phi-tree-spect}), we obtain the lowest-order correlator contributed by the scalar diquark component:
\begin{align}
\Phi^{(0)}_s(x,\bk)&\equiv \frac{N_s^2(1-x)^3}{32 \pi^3 P^+}\frac{\left[ (k\ssh +m)(P\ssh +M) (k\ssh +m)\right]}{(\bk^2+L_s^2)^4}, \label{lophis}
\end{align}
and by the axial-vector diquark component:
\begin{eqnarray}
 \Phi^{(0)}_{v}(x,\bk)
&\equiv& \frac{N_v^2(1-x)^3}{64 \pi^3 P^+}d_{\mu\nu}(P-k)\nonumber\\
&\times& \frac{\left[(k\ssh +m)\gamma^{\mu}(P\ssh -M)\gamma^{\nu} (k\ssh+m)\right]}{(\bk^2+L_v^2)^4}, \label{lophiv}
\end{eqnarray}
where $k^+ = xP^+$, and $d_{\mu\nu}$ is the polarization sum (the propagator) of the axial-vector diquark.

To calculate the T-even DF $e(x,k_T^2)$, it is sufficient to apply the lowest order results (\ref{lophis}) and (\ref{lophiv}) for the correlator.
However, (\ref{lophis}) and (\ref{lophiv}) lead to a vanishing $g^\perp (x,k_T^2)$ since it is T-odd.
To obtain the nonzero result for $g^\perp (x,k_T^2)$, one needs to consider the interference between the lowest-order amplitude $\mathcal M^{(0)}$ and the one-loop-order amplitude $\mathcal M^{(1)}$ for generating the necessary phase difference.
In the spectator model, this interference gives rise to the following contributions to the quark correlator:
\begin{align}
  \Phi_s^{(1)}
(x,\bm k_T)
&\equiv
-i e_qe_s N_{s}^2  { (1-x)^3\over 32\pi^3 P^+}{1\over (\bm{k}_T^2+L_s^2)^2}\nonumber \\
\hspace{-1cm}&\times \int {d^2 \bm q_T\over (2\pi)^2}
{ \left[(\kslash -q\ssh+m) (\Pslash+M)(\kslash +m)\right]
\over \bm q_{T}^2  ((\bm{k}_T-\bm{q}_T)^2+L_s^2)^2}
 , \label{phis1}\\
 \Phi^{(1)}_{v}
(x,\bm k_T)
&\equiv
-i e_q N_v^2  { (1-x)^2\over 128\pi^3 (P^+)^2}{1\over (\bm{k}_T^2+L_v^2)^2}\nonumber\\
&\times\int {d^2 \bm q_T\over (2\pi)^2} \,
 d_{\rho\alpha}(P-k)\, (-i\Gamma^{+,\alpha\beta}) \nonumber\\
 &\times d_{\sigma\beta}(P-k+q) \nonumber\\
&\times{ \left[(\kslash -q\ssh+m) \gamma^\sigma(\Pslash-M)\gamma^\rho (\kslash +m)\right]
\over \bm q_T^2  ((\bm{k}_T-\bm{q}_T)^2+L_v^2)^2},
\end{align}
where $q^+=0$ is understood, and
$\Gamma_s^\mu $ or $\Gamma_v^{\mu,\alpha\beta}$ is
the vertex between the gluon and the scalar diquark or the axial-vector diquark:
\begin{align}
 \Gamma_s^\mu &= ie_s (2P-2k+q)^\mu, \\
 \Gamma_v^{\mu,\alpha\beta} &=  -i e_v [(2P-2k+q)^\mu g^{\alpha\beta}-(P-k+q)^{\alpha}g^{\mu\beta}\nonumber\\
 &-(P-k)^\beta g^{\mu\alpha}]\label{Gamma},
\end{align}
here $e_{s/v}$ denotes the charge of the scalar/axial-vector diquark.

Substituting (\ref{lophis}) into (\ref{e}) and (\ref{phis1}) into (\ref{phitr2}), we obtain the contributions from the scalar diquark to $e$ and $g^\perp$:
\begin{align}
e^{s}(x,\bm k_T^2)&=\frac{1}{16\pi^3} \frac{N_s^2(1-x)^2}{(\bk^2+L_s^2)^4}\nonumber\\
& \times \left[(1-x)(xM+m)(M+m)\right.\nonumber\\
&\left.-(1+\frac{m}{M})\bk^2
-(x+\frac{m}{M})M_s^2\right],
 \label{es} \\
 g^{\perp s}(x,\bm k_T^2)
   &=-{N_s^2(1-x)^2\over(32\pi^3)}{e_s e_q\over4\pi}\,\nonumber\\
&\times \left[\frac{(1-x)\Lambda^2_s+(1+x)M_s^2 -(1-x)M^2}{
 {L_s^2 (\bm k_T^2+L_s^2 )^3}}\right]
\end{align}
We find that our expression for $e^s$ is the same as Eq.~(86) of  Ref.~\cite{Jakob:1997npa} when choosing $\alpha =2$, and the result for $g^{\perp s}$ has already been given in Ref.~\cite{wjmao:2012prd}.

To calculate the quark correlator contributed by the axial-vector diquark, one needs the form of the propagator $d_{\mu\nu}$.
Different forms for $d_{\mu\nu}$ will lead to different results.
In Ref.~\cite{Bacchetta:2008prd}, all the four forms of the propagator appearing in literature (Refs.~\cite{Bacchetta:plb578,Jakob:1997npa,Brodsky:2000ii,Gamberg:2007wm}) have been studied.
We find that only two of them give convergent results for $g^\perp$ even
if the dipolar form factor is applied.
The first form is
\begin{align}
 d_{\mu\nu}(P-k)  =& \,-g_{\mu\nu}\,+\, {(P-k)_\mu n_{-\nu}
 \,+ \,(P-k)_\nu n_{-\mu}\over(P-k)\cdot n_-}\,\nonumber\\
 & - \,{M_v^2 \over\left[(P-k)\cdot n_-\right]^2 }\,n_{-\mu} n_{-\nu} ,\label{d1}
\end{align}
which is the summation over the light-cone transverse polarizations of the axial-vector diquark~\cite{Brodsky:2000ii}, and has been applied to calculate leading-twist TMD DFs in Ref.~\cite{Bacchetta:2008prd}.
The second form is
\begin{align}
d_{\mu\nu}(P-k)  =& \,-g_{\mu\nu}.
\label{d2}
\end{align}
which was used in Ref.~\cite{Bacchetta:plb578}. Therefore, in this work, we will choose the above two forms for completeness.

\begin{table}
\begin{tabular}{|c||c|c|c|c|}
\hline
Diquark & $M_X$ (GeV) & $\Lambda_X$ (GeV) & $c_X$  \\
\hline
Scalar $s (ud)$ & 0.822 & 0.609  & 0.847  \\
\hline
Axial-vector $a (ud)$ & 1.492  & 0.716  & 1.061   \\
\hline
Axial-vector $a' (uu)$ & 0.890  & 0.376  & 0.880  \\
\hline
\end{tabular}
\caption{Values for the parameters to calculate the DFs in Set 1, taken from Ref.~\cite{Bacchetta:2008prd}, which are fixed by reproducing the parametrization of unpolarized~\cite{zeus} and longitudinally polarized~\cite{grsv01} parton distributions.}\label{Tab.1}
\end{table}

First of all, when choosing the first form of $d_{\mu\nu}$ (e.g., Eq.(\ref{d1})), we arrive at the following expressions for $e$ and $g^\perp$ from the axial-vector diquark component:
\begin{align}
e^{v}(x,\bk^2)&=\frac{1}{16\pi^3} \frac{N_v^2(1-x)^2}{(\bk^2+L_v^2)^4}\nonumber\\
&\times \left[(1-x)(M+m)(xM+m)\right.\nonumber\\
&\left.-(1+\frac{m}{M})\bk^2-(x+\frac{m}{M})M_v^2+ \frac{2m \bk^2}{M (1-x)}\right],
\label{ev1} \\
g^{\perp v}(x,\bm k_{T}^2)
  &= \frac{N_v^2(1-x)}{32\pi^3}{e_v e_q\over4\pi}
  \biggl\{\frac{1}{L_v^2(\bm k_T^2+L_v^2)^3}\,\nonumber\\
&\times\left[(1-x)(xM+m)^2+(1-x)^2 M^2-M_v^2\right.\nonumber\\
&\left.+xL_v^2\right]
-\frac{x}{(\bm k_T^2+L_v^2)^2\bm k_T^2}\ln\left({\bm k_T^2+L_v^2\over L_v^2}\right)\biggr\}.\label{gv1}
\end{align}
Secondly, we use the other form of $d_{\mu\nu}$ (e.g., Eq.(\ref{d2})) to obtain the alternative expressions for $e^{v}$ and $g^{\perp v}$:
\begin{align}
e^v(x,\bk^2)
   &=\frac{N_v^2(1-x)^2}{16\pi^3}
   \frac{1}{(\bk^2+L_v^2)^4}\,\nonumber\\
   &\times \left[(1-x)(M+m)(x M+m)-(2+\frac{m}{M})\bk^2\right.\nonumber\\
   &\left.+(1-x)(m^2+xM^2)-(2x+\frac{m}{M})M_v^2\right],\label{ev2}\\
g^{\perp v}(x,\bm k_{T}^2)
  &= \frac{N_v^2(1-x)}{32\pi^3}{e_v e_q\over4\pi}
  \biggl\{\frac{1}{L_v^2(\bm k_T^2+L_v^2)^3}\,\nonumber\\
&\times\left[(1-x)(xM+m)^2+(1-x)^2(2-x) M^2\right.\nonumber\\
&\left.-(x+2)M_v^2-(1-x)L_v^2-x \bk^2\right]\nonumber\\
&-\frac{x}{(\bm k_T^2+L_v^2)^2\bm k_T^2}\ln\left({\bm k_T^2+L_v^2\over L_v^2}\right)\biggr\}.
\label{gv2}
\end{align}

With $f^s$ and $f^v$ (here $f$ denotes an arbitrary TMD DF) at hand, one can construct the DFs for $u$ and $d$ valence quarks.
However, in doing this calculation, there is a degree of freedom one can choose, which is the relation between quark flavors and diquark types.
As shown in Ref.~\cite{Bacchetta:2008prd}, a general relation can be cast into
\begin{align}
f^u=c_s^2 f^s + c_a^2 f^a,~~~~f^d=c_{a^\prime}^2 f^{a^\prime}\label{ud},
\end{align}
here $a$ and $a^\prime$ denote the
vector isoscalar diquark $a(ud)$ and the vector isovector diquark $a(uu)$, respectively, and
$c_s$, $c_a$ and $c_{a^\prime}$ are the parameters of the model.
In Ref.~\cite{Bacchetta:2008prd}, these parameters as well as the mass parameters (such as the diquark masses $M_X$, cut-off parameters $\Lambda_X$) are fitted from the ZEUS~\cite{zeus} and GRSV01~\cite{grsv01} DF sets.
Particularly, in Ref.~\cite{Bacchetta:2008prd}, the mass parameters for different vector diquark types are treated differently, that is, the two isospin states of the vector diquark are distinguished.

Different from Eq.~(\ref{ud}),  a commonly used approach
in the previous spectator models~\cite{Bacchetta:plb578,Jakob:1997npa} to construct the distributions of $u$ and $d$ valence quarks can be expressed as follows:
\begin{align}
f^u=\frac{3}{2}f^s+\frac{1}{2} f^a,~~~~f^d=f^{a^\prime}, \label{set2}
\end{align}
here the coefficients in front of $f^X$ are obtained from the SU(4) spin-flavor symmetry of the proton wave function.
In this case, the mass parameters for different axial diquark are the same.

\begin{figure}
  \includegraphics[width=0.49\columnwidth]{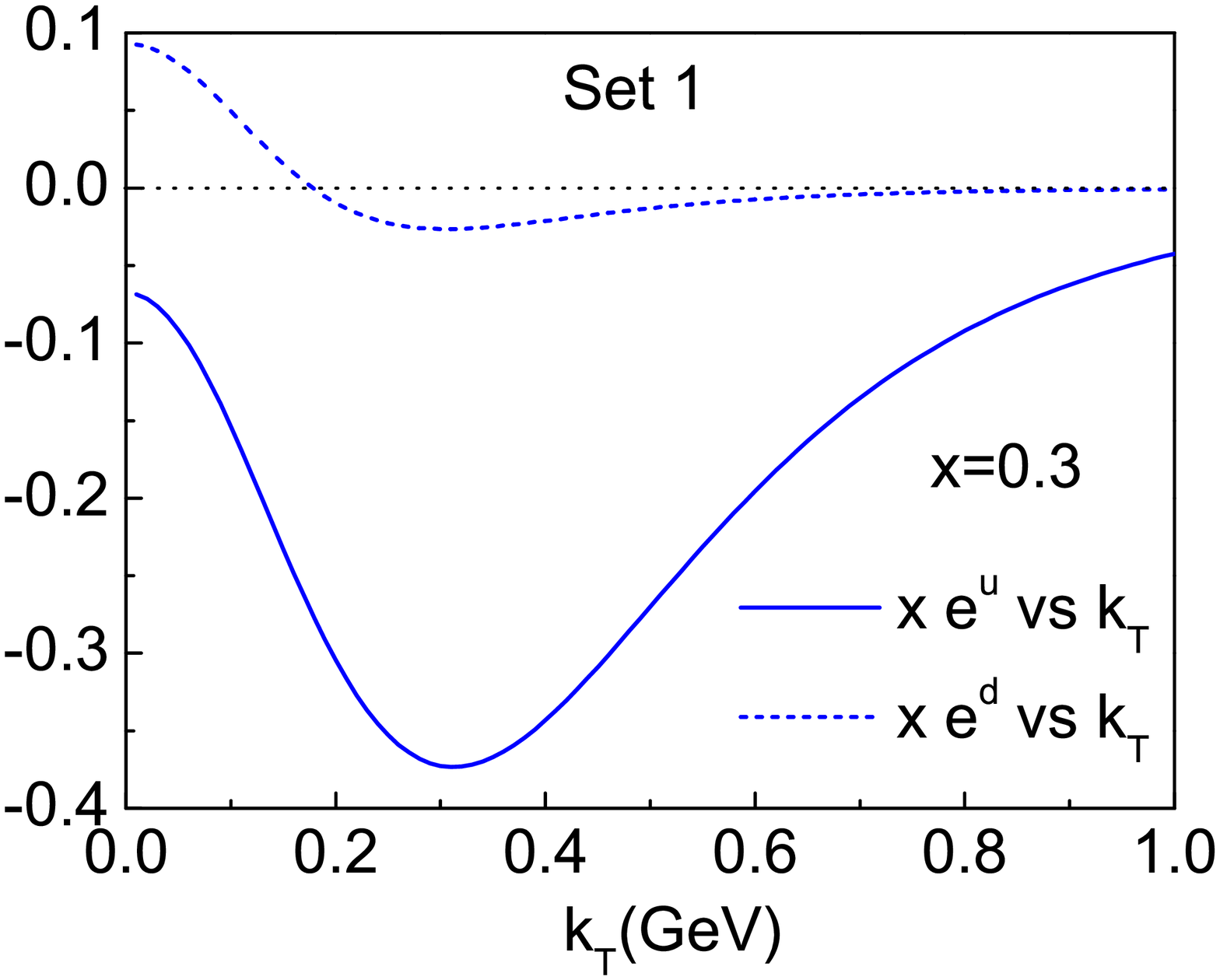}
  \includegraphics[width=0.49\columnwidth]{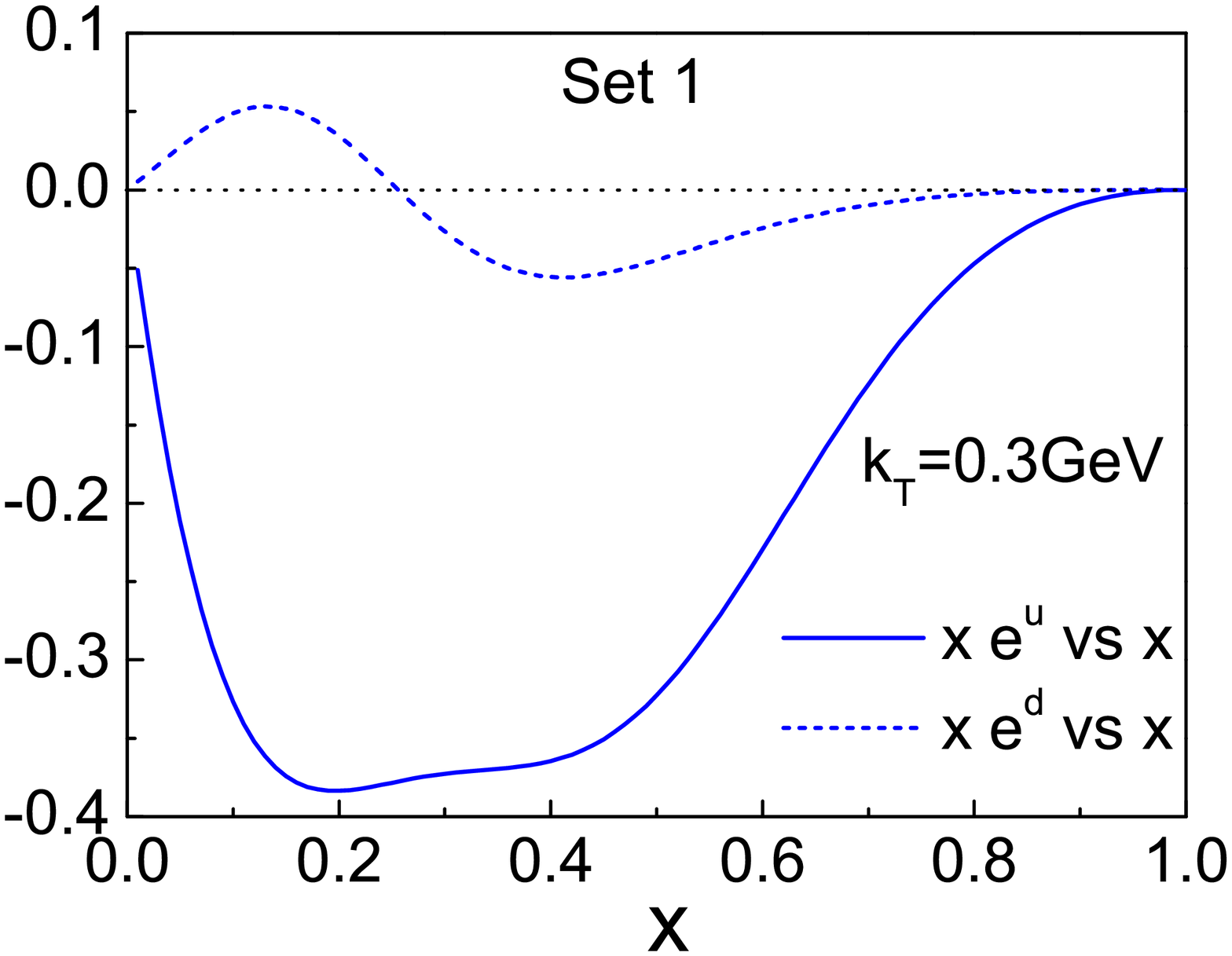}\\
  \caption{The $x$ and $k_T$ dependences of $xe(x,\bm k_T^2)$ in Set 1. Left panel: the shapes of $x e^{u}$ (solid line) and $x e^{d}$ (dashed line) as  functions of $k_T$ at $x=0.3$; right panel: the shapes of $x e^{u}$ (solid line) and $x e^{d}$ (dashed line) as functions of $x$ at  $k_T=0.3\,\text{GeV}$.}\label{edependence.1}
\end{figure}

\begin{figure}
  \includegraphics[width=0.49\columnwidth]{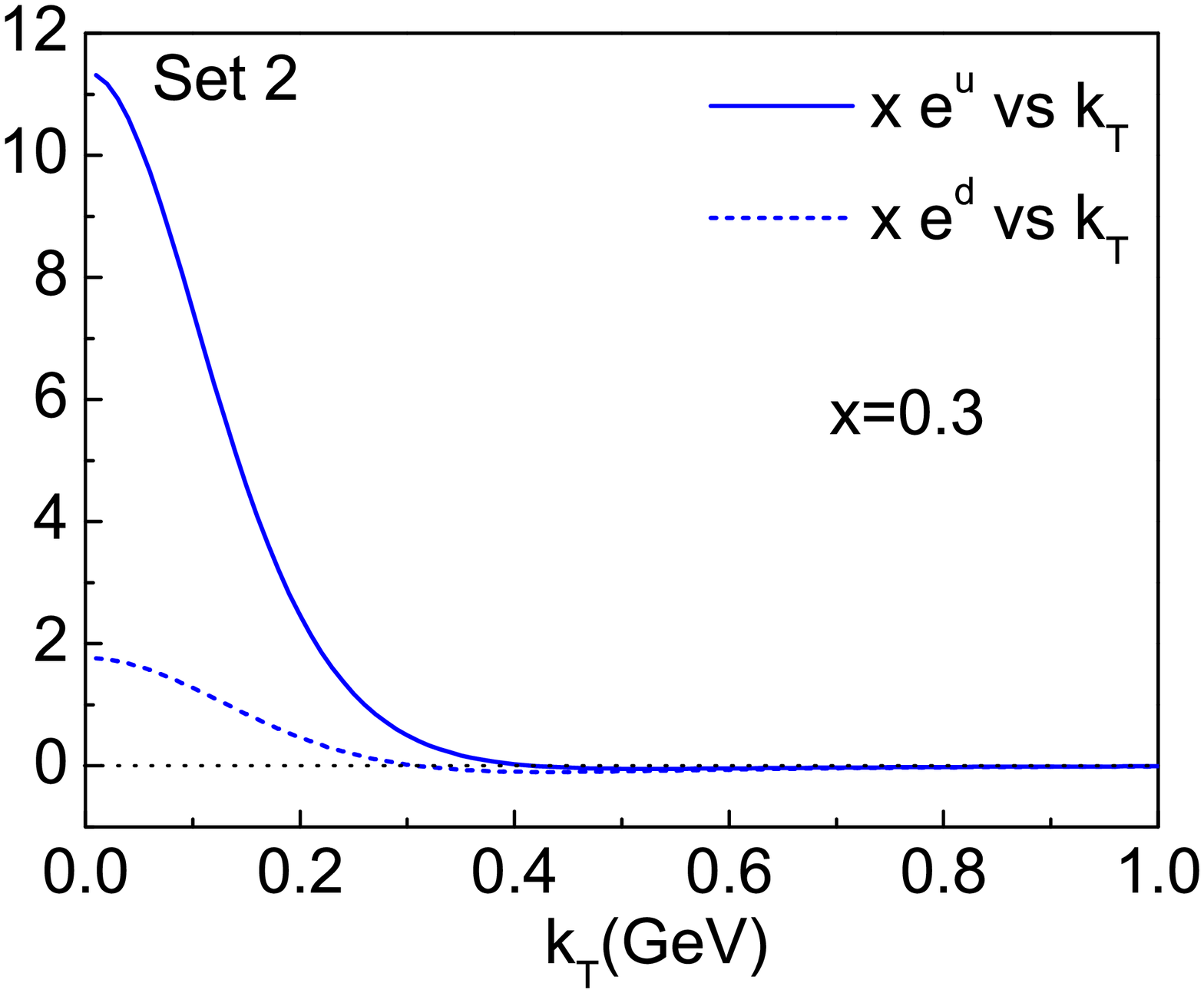}
  \includegraphics[width=0.49\columnwidth]{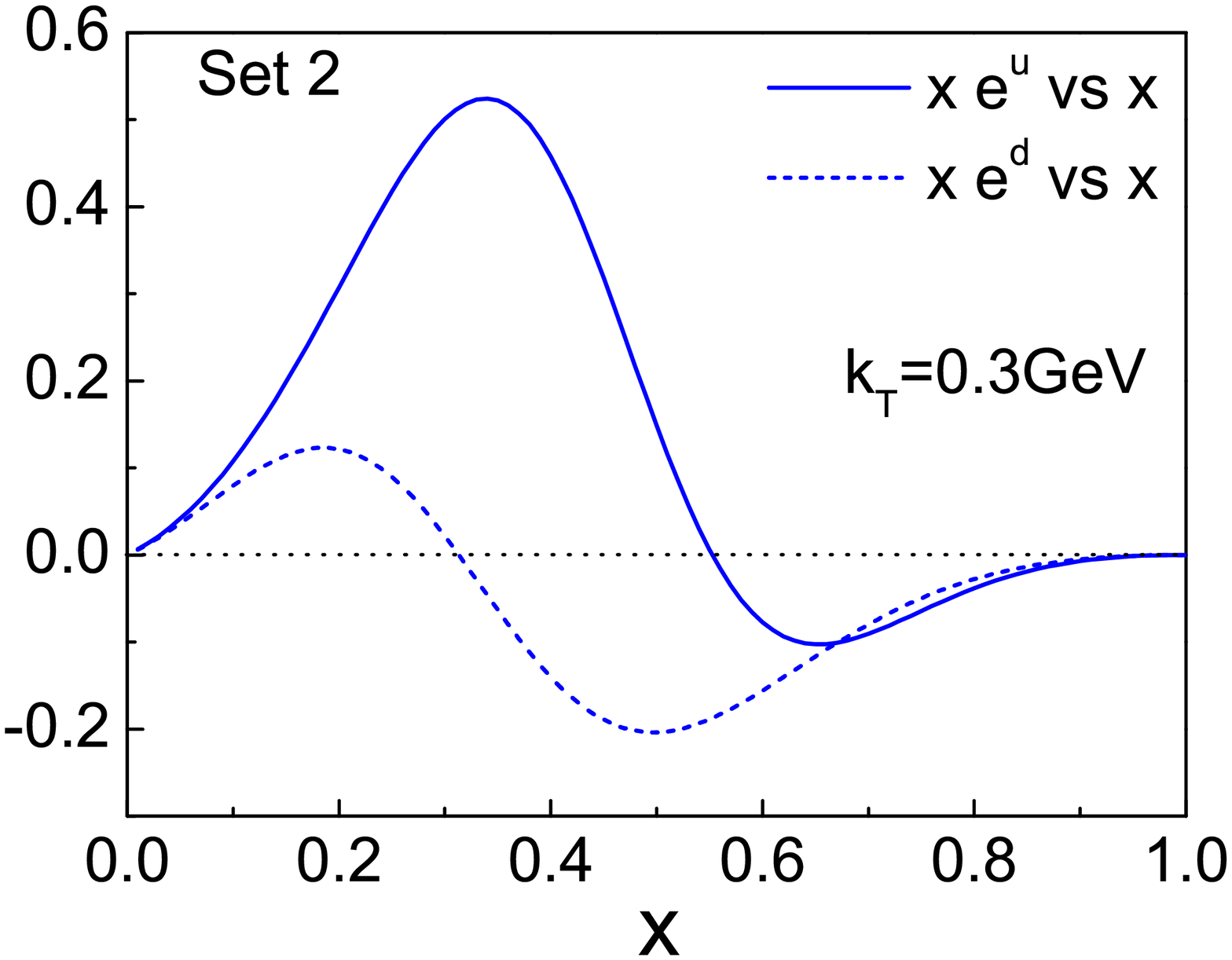}\\
  \caption{Similar to Fig.~\ref{edependence.1}, but for $e$ in Set 2.}\label{edependence.2}
\end{figure}

\begin{figure}
  \includegraphics[width=0.49\columnwidth]{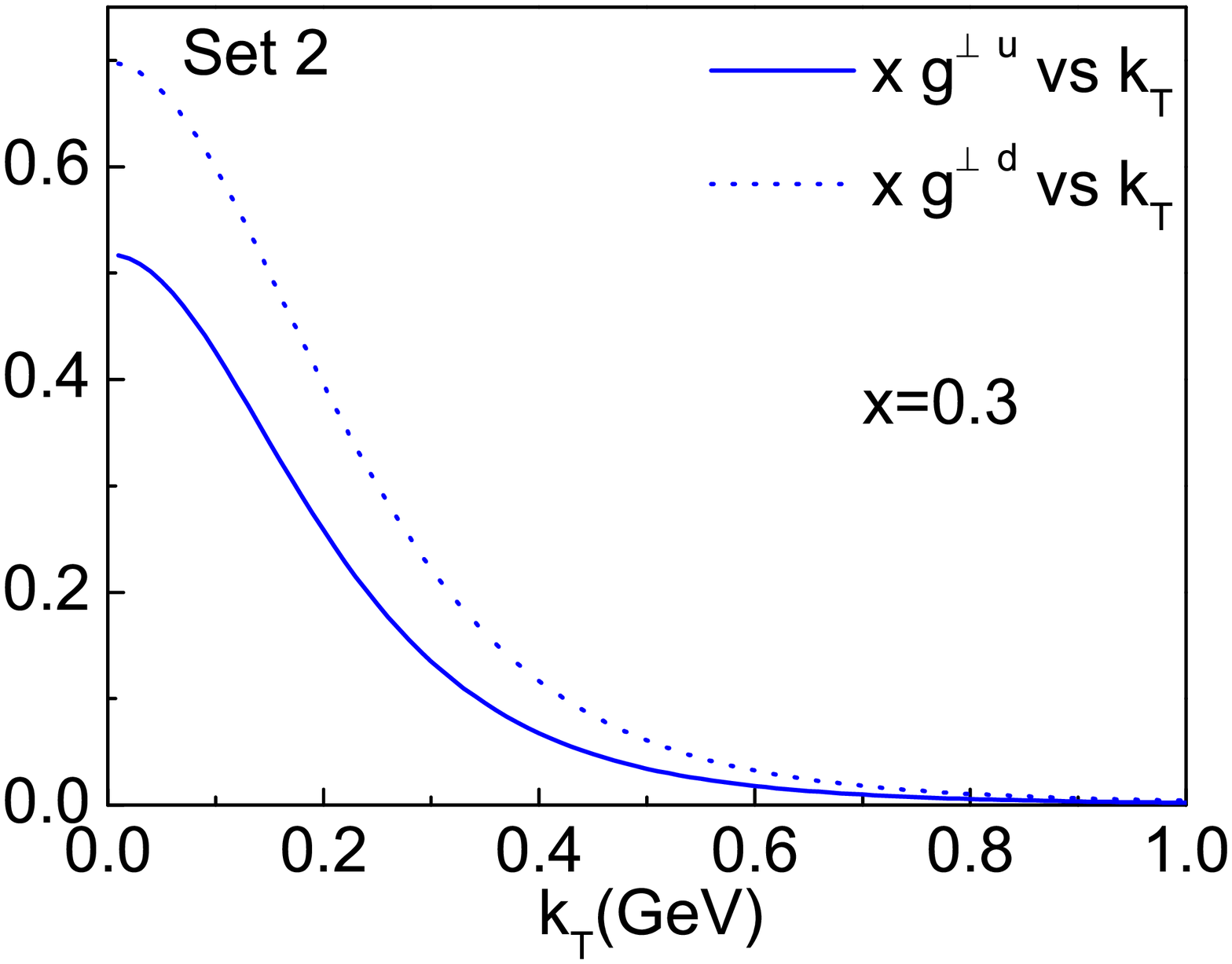}
  \includegraphics[width=0.49\columnwidth]{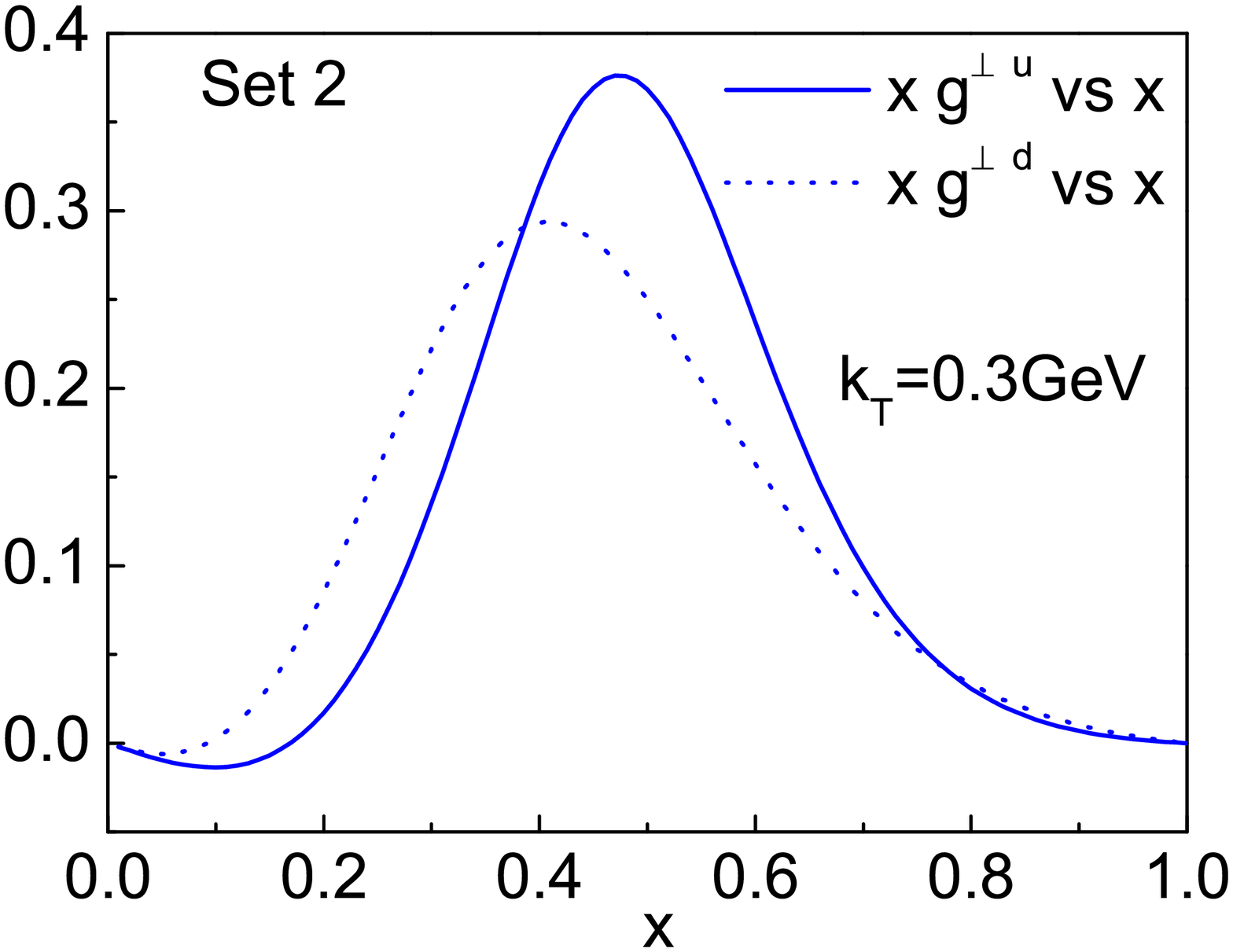}\\
  \caption{The $x$ and $k_T$ dependences of $xg^\perp(x,\bm k_t^2)$ in Set 2. Left panel: the shapes of $x g^{\perp u}$ (solid line) and $x g^{\perp d}$ (dashed line) as  functions of $k_T$ at $x=0.3$; right panel: the shapes of $x g^{\perp u}$ (solid line) and $x g^{\perp d}$ (dashed line) as functions of $x$ at  $k_T=0.3\,\text{GeV}$.}\label{gperpud}
\end{figure}

In this work, we will consider both the relations given in (\ref{ud}) and
(\ref{set2}), combined with two choices for the axial-diquark propagator, to obtain two sets of the DFs $e$ and $g^\perp$ for the
valence quarks.
To calculate the first set of $e$ and $g^\perp$ (we label them as the ``Set 1" DFs), we use $e^v$ and $g^{\perp v}$ calculated from the first choice for the propagator $d_{\mu\nu}$ (\ref{d1}) together with the relation shown in (\ref{ud}), which has also been
applied in Ref.~\cite{Bacchetta:2008prd}.
For the parameters used in this calculation, again we adopt them (listed in Table.~\ref{Tab.1}) from Ref.~\cite{Bacchetta:2008prd} for consistency.
In Fig.~\ref{edependence.1} we plot the $x$ and $k_T$ dependences of $e(x,k_T^2)$ in Set 1 for the $u$ and $d$ valence quarks.
We will not present the numerical results for $g^\perp$ since the corresponding curves have been shown in Fig.~1 of Ref.~\cite{wjmao:2012prd}.
We also apply the expressions of $e^v$ and $g^{\perp v}$ resulted from the second form of $d_{\mu\nu}$ (\ref{d2}), along with the SU(4) relation (\ref{set2}), to calculate the numerical results of $e$ and $g^\perp$ for the $u$ and $d$ valence quarks (we label them as the ``Set 2" DFs).
In Fig.~\ref{edependence.2} and Fig.~\ref{gperpud} we show the DFs $e$
and $g^\perp$ as functions of $x$ and $k_T$ in Set 2, respectively.

Figs.~\ref{edependence.1},~\ref{edependence.2},~\ref{gperpud} and Fig.~1 in Ref.~\cite{wjmao:2012prd} show that different approaches in the spectator model will lead to quite different results for $e$ and $g^\perp$,
including their flavor dependences, the sizes and signs.
In Set 1 $e^u$ is negative while that in Set 2 is mostly positive, also
the $k_T$ dependences for $e$ in the two sets are very different.
First, $e^u$ and $e^d$ in Set 2 monotonically decrease with increasing $k_T$, while those in Set 1 increase in the low $k_T$ region then start to decrease at $k_T=0.3$~GeV.
Second, at small $k_T$ the size of $e$ in Set 2 is much larger than that in Set 1.
Similarly, We find that at small $k_T$, the size of $f_1$ in Set 2 is also much  larger than that in Set 1, although the $x$ dependence of the collinear DF $f_1(x)$ in Set 1 is similar to $f_1(x)$ in Set 2.
We would like to point out that the size and sign of $e(x)$ in Set 2 are similar to the results in Ref.~\cite{Jakob:1997npa}, where another propagator for the axial-vector diquark was used:
\begin{align}
d_{\mu\nu}(P-k)=-g_{\mu\nu} +{P_\mu P_\nu\over M_v^2}.
\end{align}
Also in Set 2, the sign of $e(x)$ in the small and moderate $x$ regions is consistent with the calculation from the chiral quark-soliton model~\cite{Schweitzer:2003prd67}.
For the DF $g^\perp$, we find that its flavor dependence is different in the two sets, such that in Set 1 (see Fig.~1 in Ref.~\cite{wjmao:2012prd}) the size of $g^{\perp u}$ is several times larger than that of $g^{\perp  d}$, while in Set 2 the sizes of them are comparable.

\begin{table}
\begin{tabular}{|c||c|c|c|c|}
\hline
Diquark &$M_X$ (GeV) & $\Lambda_X$ (GeV) & $c_X^2$ \\
\hline
Scalar $s(ud)$ &0.6 &0.5 &1.5  \\
\hline
Axial-vector $a(ud)$ &0.8 &0.5 &0.5  \\
\hline
Axial-vector $a'(uu)$ &0.8 &0.5 &1.0 \\
\hline
\end{tabular}
\caption{Values for the parameters to calculate the DFs in Set 2, taken from Ref.~\cite{Bacchetta:plb578}.}
\label{Tab.2}
\end{table}

\section{Numerical results for beam single spin asymmetries in three different pion production}
\label{BSAs}

In this section, we perform our phenomenological analysis
on the beam SSAs for $\pi^+$, $\pi^-$ and $\pi^0$ at the kinematics of HERMES and CLAS.
The process under study is the SIDIS with a longitudinally polarized lepton beam:
\begin{align}
e (\ell) \, + \, p (P) \, \rightarrow \, e' (\ell')
\, + \, h (P_h) \, + \, X (P_X)\,,
\label{sidis}
\end{align}
where $\ell$ and $\ell'$ denote the momenta of the incoming and scattered electon/positron, and $P$ and $P_h$ denote those of the target nucleon and the final-state hadron. The differential cross section of the SIDIS is expressed by the invariants:
\begin{align}
&x = \frac{Q^2}{2\,P\cdot q},~~~
y = \frac{P \cdot q}{P \cdot l},~~~
z = \frac{P \cdot P_h}{P\cdot q},~~~\gamma={2Mx\over Q},~~~\nonumber\\
&Q^2=-q^2, ~~~
s=(P+\ell)^2,~~~
W^2=(P+q)^2,~~~
\end{align}
here $q=\ell-\ell'$ is the momentum of the virtual photon, and $W$ is the invariant mass of the hadronic final state.
\begin{figure}
  \includegraphics[width=0.9\columnwidth]{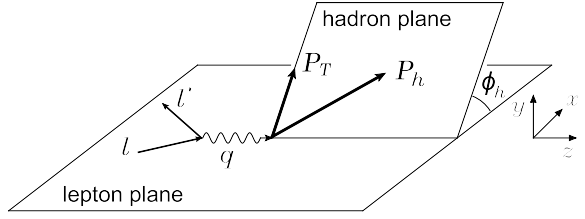}
 \caption{The kinematical configuration for the SIDIS process. The initial and scattered leptonic momenta define the lepton plane ($x-z$ plane), while the detected hadron momentum together with the $z$ axis identify the hadron production plane.}
 \label{SIDISframe}
\end{figure}
The reference frame we adopt in this work is shown in Fig.~\ref{SIDISframe}, where the virtual photon and the target proton are collinear and along the $z$ axis. Furthermore, we use $\bk$ to denote the intrinsic transverse momentum of the quark inside the proton, and use $\bP$ to denote the transverse momentum of the detected hadron. The transverse momentum of the hadron $h$ with respect to the direction of the fragmenting quark is denoted by $\bp$. The azimuthal angle between the lepton and the hadron planes is defined as $\ph$.

Generally, the differential cross section of SIDIS for a longitudinally polarized beam with helicity $\lambda_e$ off an unpolarized hadron can be expressed as ~\cite{Bacchetta:0611265}:
\begin{align}
\label{HLT}
\frac{d\sigma}{d\xbj dy\,d\zh dP^2_T d\ph} &=\frac{2\pi \alpha^2}{\xbj y Q^2}\frac{y^2}{2(1-\varepsilon)}
 \Bigl( 1+ \frac{\gamma^2}{2\xbj} \Bigr)
  \left\{ F_{UU} \right.\nonumber\\ & \left.+ \lambda_e \sqrt{2\varepsilon(1-\varepsilon)} \sin \phi_h \,\,F^{\sin \ph}_{LU}\right\},
\end{align}
where $F_{UU}$ and $F_{LU}^{\sin\phi_h}$ are the helicity-averaged and helicity-dependent structure functions, respectively. The subscripts of the above two structure functions stand for different polarizations of the beam or the target.
The ratio of the longitudinal and transverse photon flux denoted by $\varepsilon$ can be given as:
\begin{align}
\varepsilon=\frac{1-y-\gamma^2y^2/4}{1-y+y^2/2+\gamma^2y^2/4}.
\end{align}

In the parton model, based on the tree-level factorization adopted in Ref.~\cite{Bacchetta:0611265}, the two structure functions in Eq.~(\ref{HLT}) can be expressed as the convolutions of twist-2 and twist-3 TMD DFs and FFs.
With the help of the notation
\begin{align}
\mathcal{C}[w fD] &=x\sum_q e_q^2\int d^2\bm k_T\int d^2 \bm p_T\delta^2(z\bm k_T-\bm P_T+\bm p_T) \nonumber\\
&\times w(\bm k_T, \bm p_T)f^q(x,\bm k_T^2) D^q(z,\bm p_T^2),
\end{align}
and the reference frame we choose, $F_{UU}$ and $F_{LU}^{\sin\phi_h}$ are given by the following expressions~\cite{Bacchetta:0611265}:
\begin{align}
F_{UU} & = \mathcal{C}[f_1 D_1], \label{FUU}\\
F^{\sin \ph}_{LU} & =  \frac{2M}{Q} \,
\mathcal{C}\,
   \left[\frac{\boldsymbol{\hat{P}_{T}} \cdot \boldsymbol{p_T}}{z M_h}
         \left(\frac{M_h}{M}\,f_1\, \frac{\tilde{G^{\perp}}}{z} + \xbj\, e H_1^{\perp} \right)\right.\nonumber\\
&\left.+\frac{\boldsymbol{\hat{P}_{T}}\cdot
    \boldsymbol{k_T}}{M}\left(\frac{M_h}{M}\, h_1^{\perp} \frac{\tilde{E}}{z} +\xbj\, g^{\perp} D_1\right)\right] ,\label{FLU}
\end{align}
where $\hat {\bm P}_T= {\bP\over P_T}$ with $P_T =|\bP|$, and $M_h$ is the mass of the final-state hadron.
The beam SSA $A_{LU}^{\sin\phi}$ as a function of $P_T$ therefore can be expressed as
\begin{align}
A_{LU}^{\sin\phi_h}(P_T) &= \frac{\int dx \int dy \int dz \;\frac{1}{x y Q^2}\frac{y^2}{2(1-\varepsilon)}}{\int dx \int dy \int dz \;\frac{1}{x y Q^2}\frac{y^2}{2(1-\varepsilon)}}\nonumber\\
&\times \frac{
\Bigl( 1+ \frac{\gamma^2}{2x} \Bigr) \sqrt{2\varepsilon(1-\varepsilon)} \;F_{LU}^{\sin\phi_h}}{\Bigl( 1+ \frac{\gamma^2}{2x} \Bigr) \;F_{UU}}. \label{asy}
\end{align}
The $x$-dependent and the $z$-dependent asymmetries can be defined in a similar way.

Eq.~(\ref{asy}) shows that it is the structure function $F^{\sin \ph}_{LU}$ that gives rise to the $\sin\phi_h$ beam SSAs.
As we can see from Eq.~(\ref{FLU}), $F^{\sin \ph}_{LU}$ receives various contributions from the convolutions of the twist-3 TMD DFs and FFs with the twist-2 ones.
In the following calculation, we will
neglect the contributions from the quark-gluon-quark correlators (often referred to as the Wandzura-Wilczek approximation~\cite{Wandzura:1977qf}), which is equivalent to setting all the functions with a tilde to zero.
It is worthwhile to point out that a calculation from the spectator model~\cite{Gamberg:2008yt} as well as a model-independent analysis~\cite{Metz:2008prl} on the $T$-odd quark-gluon-quark
correlators shows that the gluonic (partonic) pole contributions for FFs vanish.
The FF $\tilde{G}^\perp(x,\bm p_T^2)$ appears in the decomposition of the $T$-odd part of the TMD quark-gluon-quark correlator~\cite{Bacchetta:0611265,Boer:2003cm}, for which the gluonic pole contribution should play an essential role.
Whether the vanishing gluonic pole matrix elements for collinear FFs can be generalized to the case of TMD FFs deserves further study~\cite{Gamberg:2010uw}.
Nevertheless, we ignore the $\tilde{G}^\perp$ and $\tilde{E}$ contributions based on the Wandzura-Wilczek approximation.

According to the above arguments, there are two remaining terms that give contributions to the structure function $F^{\sin \ph}_{LU}$.
One is the Collins-effect term $e H_1^\perp$, which has been applied to analyze the beam SSA of $\pi^+$ production in Refs.~\cite{Gamberg:2003pz,Efremov:2002ut}.
The other is the $g^\perp D_1$ term that was proposed in Ref.~\cite{Bacchetta:g2004plb}, and was adopted to calculate the beam
SSA of $\pi^0$ production~\cite{wjmao:2012prd} recently.
Thus, in the following calculation of
the beam SSAs for $\pi^+$, $\pi^-$ and $\pi^0$, we will take both terms into consideration and arrive at
\begin{align}
\label{FLUa}
   F^{\sin \ph}_{LU} &\approx   \frac{2Mx}{Q} \,
  \sum_{q=u,d} e_q^2 \int d^2 \! \bk \biggl\{ \frac{\hat{\bm P}_T \cdot (\bP-z\bk)}{z M_h}\,\nonumber\\
  &\times \left[x\, e^q(x,\bk^2) H_1^{\perp q}\left(z,(\bP-z\bk)^2\right)\right]\,\nonumber\\
   &+\frac{\hat{\bm P}_T\cdot\bk} {M}
         \left[x\, g^{\perp q}(x,\bk^2) D_1^{q}\left(z,(\bP-z\bk)^2\right)\right]  \biggr\}\,.
\end{align}

For the twist-3 TMD DFs $e$ and $g^\perp$, we apply the results obtained in the previous section.
As for the Collins FF $H_1^{\perp}$ for different pions, we adopt the following relations:
\begin{align}
 H_1^{\perp \pi^+/u}&=H_1^{\perp \pi^-/d}\equiv H_{1 fav}^{\perp} ,\\
 H_1^{\perp \pi^+/d}&=H_1^{\perp \pi^-/u}\equiv H_{1 unf}^{\perp} ,\\
 H_1^{\perp \pi^0/u}&=H_1^{\perp \pi^0/d}\equiv{1\over 2}\left( H_{1 fav}^{\perp}+H_{1 unf}^{\perp}\right),
\end{align}
where $H_{1 fav}^{\perp}$ and $H_{1 unf}^{\perp}$ are the favored and unfavored Collins functions, for which we use the parameterized results from Ref.~\cite{Anselmino:2008jk}.
\begin{figure}
  \includegraphics[width=0.9\columnwidth]{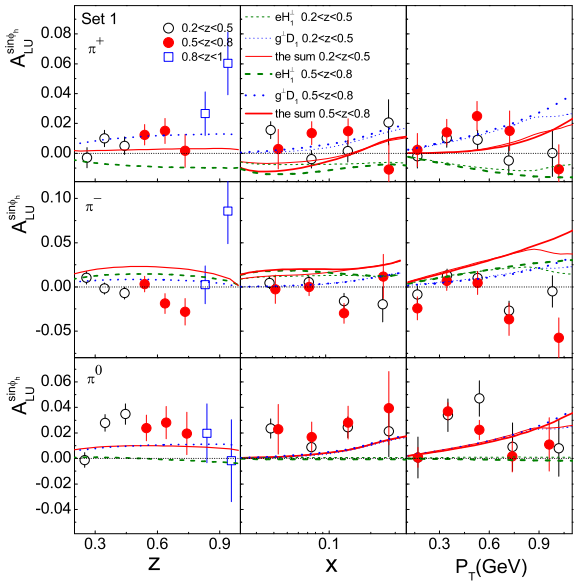}
\caption{The beam SSAs $A_{LU}^{\sin\phi_h}$ for $\pi^+$, $\pi^-$  and $\pi^0$ production in SIDIS at HERMES calculated from the Set 1 twist-3 DFs. The dashed, dotted and solid curves show the results from the Collins effect term, the $g^\perp D_1$ term and the total contribution, respectively. In the central and right panels, the thin and the thick lines correspond to the results for the low-$z$ ($0.2<z<0.5$) and mid-$z$ ($0.5<z<0.8$) regions. The data are from Ref.~\cite{hermes07}, with open circles, full circles, and open squares for $0.2<z<0.5$, $0.5<z<0.8$, and $0.8<z<1$ regions. The error bars represent the statistical uncertainties.}
  \label{HERMESa1}
\end{figure}

For the TMD FF $D_1^q\left(z,\bp^2\right)$ that couples with the distribution $g^\perp$, we assume its $p_T$ dependence  has a Gaussian form
\begin{align}
D_1^q\left(z,\bp^2\right)=D_1^q(z)\, \frac{1}{\pi \langle p_T^2\rangle}
\, e^{-\bm p_T^2/\langle p_T^2\rangle},
\end{align}
where $\langle p_T^2\rangle$ is the Gaussian width for $p_T^2$.
Following the fitted result in Ref.~\cite{Anselmino:2005prd}, we choose $\langle p_T^2\rangle=0.2$ \textrm{GeV}$^2$ in the calculation.
For the integrated FFs $D_1^q(z)$, we will adopt the Kretzer parametrization~\cite{kretzer:2000prd}.
Finally, throughout the paper, we consider the following kinematical constraints~\cite{Boglione:2011} on the intrinsic transverse momentum of the initial quarks in our calculation:
\begin{equation}
 \begin{cases}
k_{T}^2\leq(2-x)(1-x)Q^2, ~~~\textrm{for}~~0< x< 1 
; \\
k_{T}^2\leq \frac{x(1-x)} {(1-2x)^2}\, Q^2, ~~~~~~~~~~~~\textrm{for}~~x< 0.5.
\end{cases}\label{constraints}
 \end{equation}
The first constraint in Eq.~(\ref{constraints}) is obtained by requiring the energy of the parton to be less than the energy of the parent hadron, while the second constraint
arises from the requirement that
the parton should move in the forward direction with respect to the parent hadron~\cite{Boglione:2011}. For the region $x<0.5$, there are two upper limits for $k_T^2$ at the same time; it is understood that the smaller one should be chosen.

\begin{figure}
  \includegraphics[width=0.9\columnwidth]{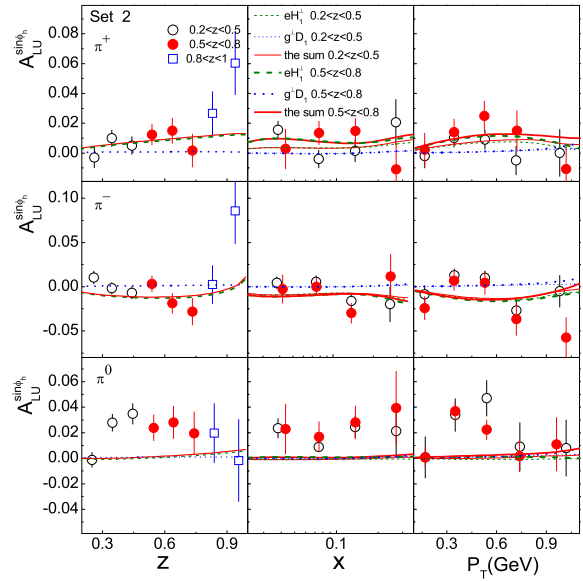}
  \caption{Similar to Fig.~\ref{HERMESa1}, but calculated from the twist-3 DFs in Set 2.}
  \label{HERMESb4}
\end{figure}

To perform numerical calculation on the beam SSAs of pion production in SIDIS at HERMES, we adopt the following kinematical cuts~\cite{hermes07}:
\begin{align}
&0.023 < x < 0.4,\,0 < y < 0.85, \,1 \textrm{GeV}^2< Q^2 < 15\, \textrm{GeV}^2, \nonumber\\
& W^2 > 4\, \textrm{GeV}^2,~~
~~ 2\,\textrm{GeV} < E_h < 15\, \textrm{GeV},
\end{align}
where $E_h$ is the energy of the detected final-state hadron in the
target rest frame.
In the left, central, and right panels of Fig.~\ref{HERMESa1} and Fig.~\ref{HERMESb4}, we show the results of the beam SSAs for $\pi^+$, $\pi^-$ and $\pi^0$ as functions of $z$, $x$, and $P_T$ and compare them with the HERMES data~\cite{hermes07}. Fig.~\ref{HERMESa1} shows the results when using the TMD DFs $e$ and $g^\perp$ in Set 1, while Fig.~\ref{HERMESb4} shows the results calculated from Set 2.

Comparing the theoretical curves with the data, one can see that in the case of the first set, our results can describe the data for $\pi^+$ and $\pi^0$, as shown in Fig.~\ref{HERMESa1}; while our calculation miss the sign of the asymmetry for $\pi^-$.
After carefully examining different contributions to the beam SSA at HERMES, we find that for $\pi^+$ production, the $e\,H_1^\perp$ term and the $g^\perp\, D_1$ term give rise to the asymmetries with opposite signs, while for $\pi^-$ production, both terms contribute positive asymmetries.
In the case of the second set, our results agree with the data for $\pi^+$ and $\pi^-$ fairly well, on the contrary, the prediction for $\pi^0$ asymmetry largely underestimates the data.
We also find that, in this case, the leading contribution is from the $e\,H_1^\perp$ term, however, the contribution from the $g^\perp D_1$ term is almost negligible.
This is due to the fact that in this set the size of $g^\perp$ at small $k_T$ is much smaller than that of $e$.
In both sets, the contribution from the $e H_1^\perp$ term to the $\pi^0$ asymmetry is very small.
This is an expected result since the favored and unfavored Collins functions have almost similar sizes but opposite signs.

Furthermore, to make a thorough comparison, we also make the prediction at CLAS, where the asymmetries for $\pi^+$, $\pi^-$ and $\pi^0$ are being measured by using a $5.5\,\text{GeV}$ longitudinally polarized electron beam off the proton target~\cite{Avakian13}.
Again, we adopt the kinematical constraints on $k_T$ in Eq.~(\ref{constraints}) and apply the following kinematical cuts to perform the numerical calculation:
\begin{align}
&0.1<x<0.6,~~ 0.4<z<0.7,~~ Q^2>1\, \textrm{GeV}^2,\nonumber\\
&P_T>0.05\,\textrm{GeV},~~ W^2>4\,\textrm{GeV}^2.
\end{align}
\begin{figure}
  \includegraphics[width=0.9\columnwidth]{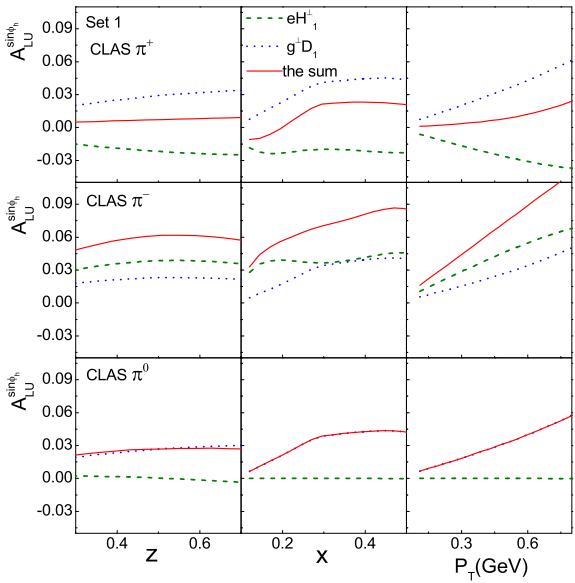}
  \caption{The beam SSA $A_{LU}^{\sin\phi_h}$ for $\pi^+$, $\pi^-$ and $\pi^0$  in SIDIS at CLAS, calculated from the twist-3 DFs in Set 1.  The dashed, dotted and solid curves show the asymmetries from the $e H_1^\perp$ term, the $g^\perp D_1$ term and the sum of the two terms. }
  \label{clasa1pi}
\end{figure}

\begin{figure}
  \includegraphics[width=0.9\columnwidth]{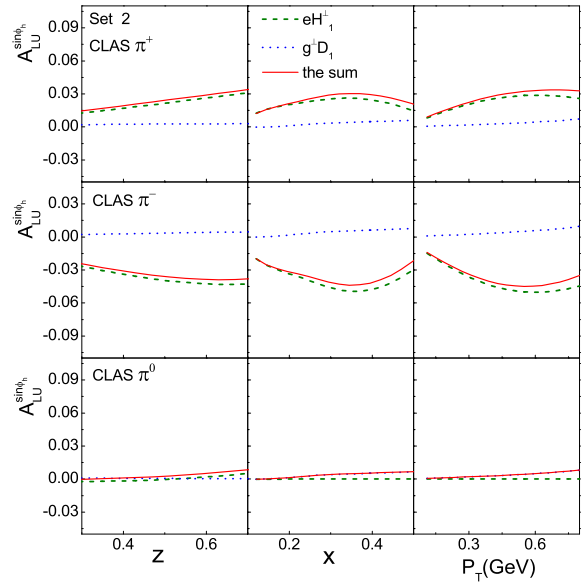}
  \caption{Similar to Fig.~\ref{clasa1pi}, but calculated from the twist-3 DFs in Set 2.}
  \label{clasb4pi}
\end{figure}

In the left, central and right panels of Fig.~\ref{clasa1pi}, we plot the $z$, $x$, and $P_T$ dependences of the beam SSAs for $\pi^+$, $\pi^-$ and $\pi^0$ production calculated from the twist-3 distributions in Set 1; while in Fig.~\ref{clasb4pi}, we plot the similar asymmetries, but from the twist-3 distributions in Set 2. To distinguish different origins of the contributions, we use the dashed and dotted curves to specify the contributions from Collins effect and the $g^\perp D_1$ term. The solid curves represent the sum of the contributions from the above two terms.

It is found that two different sets of the twist-3 TMD DFs calculated in the previous section will lead to quite different beam SSAs at CLAS, including the sizes and the signs. For example, in the first set the asymmetry from the $eH_1^\perp$ term is negative for $\pi^+$ and positive for $\pi^-$; on the contrary, in the second set the result is positive for $\pi^+$ and negative for $\pi^-$.
It is also worthwhile to mention that in Set 1 the asymmetry contributed by the $g^\perp D_1$ term is dominant for $\pi^+$ and $\pi^0$; while in Set 2 the contribution from $g^\perp D_1$ for three pions is nonzero only in the large $x$ and $P_T$ regions.
Further studies are needed in order to distinguish the different contributions to the beam SSAs.

\section{conclusion}
\label{conclusion}

In this work, we investigated the beam SSAs of $\pi^+$, $\pi^-$ and $\pi^0$ production in SIDIS process.
We considered two different contributions to the beam SSAs, namely, the Collins effect and the $g^\perp D_1$ term.
By using two different choices for the propagator of the axial-vector diquark, together with different relations between the quark flavors and the diquark types,
we obtained two sets of the twist-3 TMD DFs $e$ and $g^\perp$ in the spectator model.
We find that different approaches will lead to quite different TMD DFs, including their flavor dependences, sizes and signs.
First, $e^{u}$ is negative in Set 1 but is positive in Set 2 for all $x$ and $k_T$ regions;
while $e^{d}$ is positive in the small $x$ region and turns out
to be negative in the large $x$ region.
Second, the sizes of $e$ and $f_1$ at small $k_T$ are very different in two different sets.
Third, $g^{\perp u}$ dominates over $g^{\perp d}$ in Set 1 while they are comparable in Set 2.

Using the model results for $e$ and $g^\perp$, we calculated the beam SSAs $A_{LU}^{\sin\phi_h}$ for the electroproduction of charged and neutral pions in SIDIS at the kinematics of HERMES and
CLAS, respectively.
Comparing the theoretical curves with the data measured by HERMES at 27.6 GeV, we find that from the prediction of the Set 1 DFs, our results can describe the data for $\pi^+$ and $\pi^0$, while the calculation miss the sign of the asymmetry for $\pi^-$; on the contrary, in the case of Set 2,
our results agree with the data for $\pi^+$ and $\pi^-$ fairly well, but the prediction for $\pi^0$ largely underestimates the data.
Similarly, we find that two different sets of DFs lead to different asymmetries at CLAS with the beam energy $E_e=5.5~\textrm{GeV}$.
Also, the roles of the $e H_1^\perp$ and $g^\perp D_1$ terms are different in two different sets.
In conclusion, although the spectator model calculations can describe the asymmetries for certain pion production in some kinematic regions, it seems that it is difficult to explain the asymmetries for all three pions in a consistent way from the current spectator models.
Further studies are needed to arrive at a complete description on the beam SSAs for the charged and neutral pions based on the TMD framework.

\section*{Acknowledgements}
This work is partially supported by National Natural Science
Foundation of China with Grant No.~11005018,
by SRF for ROCS from SEM, and by the Fundamental Research
Funds for the Central Universities. W. Mao is supported by the Research and Innovation Project for College Postgraduate of Jiangsu Province with Grant No. CXZZ13$\_$0079.


\begin{thebibliography}{99}

\bibitem{bdr}
 V.~Barone,
A.~Drago, and P.~G.~Ratcliffe, Phys. Rep. \textbf{359}, 1 (2002).
\bibitem{D'Alesio:2007jt}
  U.~D'Alesio and F.~Murgia,
  Prog. Part. Nucl. Phys.  \textbf{61}, 394 (2008).
\bibitem{Barone:2010ef}
  V.~Barone, F.~Bradamante, and A.~Martin,
  Prog. Part. Nucl. Phys.  \textbf{65}, 267 (2010).
\bibitem{Boer:2011fh}
  D.~Boer, M.~Diehl, R.~Milner, R.~Venugopalan, W.~Vogelsang, A.~Accardi, E.~Aschenauer, and M.~Burkardt \textit{et al.},
arXiv:1108.1713.
\bibitem{hermes00}
A. Airapetian {\it et al.} (HERMES Collaboration), Phys. Rev. Lett. {\bf 84}, 4047 (2000).
\bibitem{hermes01}
A. Airapetian {\it et al.} (HERMES Collaboration), Phys. Rev. D {\bf 64}, 097101 (2001).
\bibitem{hermes03}
A. Airapetian {\it et al.} (HERMES Collaboration), Phys. Lett. B {\bf 562}, (2003).
\bibitem{hermes05}
A. Airapetian {\it et al.} (HERMES Collaboration), Phys. Rev. Lett. {\bf 94}, 012002 (2005).
\bibitem{hermes07}
A. Airapetian {\it et al.} (HERMES Collaboration), Phys. Lett. B {\bf 648}, 164 (2007).
\bibitem{hermes09}
A. Airapetian {\it et al.} (HERMES Collaboration), Phys. Rev. Lett. {\bf 103}, 152002 (2009) .
\bibitem{hermes10}
A. Airapetian {\it et al.} (HERMES Collaboration), Phys. Lett. B {\bf 693}, 11 (2010).

\bibitem{compass05} V. Y. Alexakhin {\it et al.} (COMPASS Collaboration), Phys. Rev. Lett. {\bf 94}, 202002 (2005).
\bibitem{compass06}  E.~S.~Ageev \textit{et al.}  (COMPASS Collaboration),
  Nucl. Phys. \textbf{B765}, 31 (2007).
\bibitem{compass10} M. G. Alekseev {\it et al.} (COMPASS Collaboration), Phys. Lett. B {\bf 692}, 240 (2010).

\bibitem{Adolph:2012sn}
  C.~Adolph {\it et al.}  (COMPASS Collaboration),
  Phys.\ Lett.\ B {\bf 717} 376 (2012).


\bibitem{Adolph:2012sp}
  C.~Adolph {\it et al.}  (COMPASS Collaboration),
  Phys.\ Lett.\ B {\bf 717} 383 (2012).


\bibitem{clas04}
H. Avakian {\it et al.} (CLAS Collaboration), Phys. Rev. D {\bf 69} 112004 (2004).
\bibitem{clas10}
H. Avakian {\it et al.} (CLAS Collaboration), Phys. Rev. Lett. {\bf 105}, 262002 (2010).
\bibitem{Qian:2011py}
X.~Qian \textit{et al.} (The Jefferson Lab Hall A Collaboration),
  Phys. Rev. Lett.  \textbf{107}, 072003 (2011).
\bibitem{Aghasyan:2011ha}
  M.~Aghasyan {\it et al.},
  Phys.\ Lett.\ B {\bf 704}, 397 (2011).
\bibitem{Aghasyan:2011zz}
  M.~Aghasyan,
  AIP Conf.\ Proc.\  {\bf 1418}, 79 (2011).
\bibitem{Yuan:2004plb}
F. Yuan, Phys. Lett. B \textbf{589}, 28 (2004).
\bibitem{Boer:1998prd}
D. Boer and P. J. Mulders, Phys. Rev. D {\bf 57}, 5780 (1998).
\bibitem{Gamberg:2003pz}
  L.~P.~Gamberg, D.~S.~Hwang, and K.~A.~Oganessyan,
  Phys.\ Lett.\ B {\bf 584}, 276 (2004).
\bibitem{Efremov:2002ut}
A.~V.~Efremov, K.~Goeke, and P.~Schweitzer, Phys.\ Rev.\ D {\bf 67}, 114014 (2003).

\bibitem{Jaffe:prl67}
R. L. Jaffe and X. D. Ji, Phys. Rev. Lett. \textbf{67} (1991) 552.
\bibitem{Jaffe:npb375}
R. L. Jaffe and X. D. Ji, Nucl. Phys. B \textbf{375} (1992) 527.
\bibitem{Collins:1993npb}
J. C. Collins, Nucl. Phys. {\bf B396}, 161 (1993).
\bibitem{Afanasev:2003ze}
A.~Afanasev and C.~E.~Carlson,
 arXiv:0308163.
\bibitem{Metz:2004epja22}
A. Metz and M. Schlegel, Eur. Phys. J. A {\bf 22}, 489 (2004).
\bibitem{Bacchetta:g2004plb}
A. Bacchetta, P.J. Mulders, and F. Pijlman, Phys. Lett. B \textbf{595}, 309 (2004).
\bibitem{Brodsky:2002plb}
S.~J.~Brodsky, D.~S.~Hwang, and I.~Schmidt, Phys. Lett. B \textbf{530}, 99 (2002); Nucl. Phys. \textbf{B642}, 344 (2002).
\bibitem{Collins:2002plb}
J. C. Collins, Phys. Lett. B \textbf{536}, 43 (2002).
\bibitem{jy02}
X.~Ji and F.~Yuan, Phys. Lett. B \textbf{543}, 66
(2002).
\bibitem{Sivers:1991prd}
D. W. Sivers, Phys. Rev. D {\bf 43}, 261 (1991).
\bibitem{Anselmino:1998plb}
 M. Anselmino and F. Murgia, Phys. Lett. B \textbf{442}, 470 (1998).
\bibitem{Afanasev:2006prd74}
A. V. Afanasev and C. E. Carlson, Phys. Rev. D  \textbf{74}, 114027 (2006).
\bibitem{Gamberg:2006plb}
  L.~P.~Gamberg, D.~S.~Hwang, A.~Metz, and M.~Schlegel,
  Phys.\ Lett.\  B {\bf 639}, 508 (2006).
\bibitem{Lu:2012plb}
Z. Lu and I. Schmidt, Phys. Lett. B {\bf 712}, 451 (2012).
\bibitem{wjmao:2012prd}
W. Mao, Z. Lu, Phys. Rev. D \textbf{87}, 014012 (2013).
\bibitem{Bacchetta:2008prd}
A. Bacchetta, F. Conti, and M. Radici, Phys. Rev. D \textbf{78}, 074010 (2008).

\bibitem{Bacchetta:plb578}
A. Bacchetta, A. Sch\"afer, and J.J Yang, Phys. Lett. B {\bf 578}, 109 (2004).
\bibitem{Jakob:1997npa}
R. Jakob, P. J. Mulders, and J. Rodrigues, Nucl. Phys. {\bf A626}, 937 (1997).
\bibitem{Schweitzer:2003prd67}
P. Schweitzer, Phys. Rev. D \textbf{67}, 114010 (2003).
\bibitem{Avakian:2010prd81}
H. Avakian, A.V. Efremov, P. Schweitzer, and F. Yuan, Phys. Rev. D \textbf{81}, 074035 (2010).
\bibitem{Bacchetta:0611265}
A. Bacchetta, M.~Diehl, K.~Goeke, A.~Metz, P.~J.~Mulders, and M.~Schlegel, J.~High Energy Phys. 02 {\bf 2007} 093.
\bibitem{Brodsky:2000ii}
  S.~J.~Brodsky, D.~S.~Hwang, B.~-Q.~Ma, and I.~Schmidt,
  Nucl.\ Phys.\ {\bf B593}, 311 (2001).
\bibitem{Gamberg:2007wm}
L.~P.~Gamberg, G.~R.~Goldstein, and M.~Schlegel,
  Phys.\ Rev.\ D {\bf 77}, 094016 (2008).
\bibitem{zeus}
S. Chekanov {\it et al.}  (ZEUS Collaboration), Phys. Rev. D {\bf 67}, 012007 (2003).
\bibitem{grsv01}
M. Gl\"uck, E. Reya, M. Stratmann, and W. Vogelsang, Phys. Rev. D {\bf 63}, 094005 (2001).
\bibitem{Wandzura:1977qf}
  S.~Wandzura and F.~Wilczek,
  Phys.\ Lett.\  B {\bf 72}, 195 (1977).
\bibitem{Gamberg:2008yt}
  L.~P.~Gamberg, A.~Mukherjee, and P.~J.~Mulders,
  Phys.\ Rev.\ D {\bf 77}, 114026 (2008).
\bibitem{Metz:2008prl}
S. Meissner, A.~Metz, Phys. Rev. Lett. {\bf 102}, 172003 (2009).
\bibitem{Boer:2003cm}
  D.~Boer, P.~J.~Mulders, and F.~Pijlman,
  Nucl. Phys. {\bf B667},  201 (2003).
\bibitem{Gamberg:2010uw}
  L.~P.~Gamberg, A.~Mukherjee and P.~J.~Mulders,
  Phys.\ Rev.\ D {\bf 83}, 071503 (2011).
\bibitem{Anselmino:2008jk}
  M.~Anselmino, M.~Boglione, U.~D'Alesio, A.~Kotzinian, F.~Murgia, A.~Prokudin, and S.~Melis,
  Nucl.\ Phys.\ Proc.\ Suppl.\  {\bf 191}, 98 (2009).
\bibitem{Anselmino:2005prd}
  M. Anselmino, M.~Boglione, U.~D'Alesio, A.~Kotzinian, F.~Murgia, and A.~Prokudin, Phys. Rev. D {\bf 71}, 074006 (2005).
\bibitem{kretzer:2000prd}
S.Kretzer, Phys. Rev. D {\bf  62}, 054001 (2000).
\bibitem{Boglione:2011}
M. Boglione, S. Melis, and A. Prokudin,
Phys.\ Rev.\ D {\bf 84}, 034033 (2011).

\bibitem{Avakian13}
  H.~Avakian, to appear in proceedings of the 20th International Symposium on Spin Physics, Dubna, Russia.

  \end{thebibliography}
\end{document}